\documentclass{article}

\usepackage{arxiv}

\usepackage[utf8]{inputenc}
\usepackage[T1]{fontenc}
\usepackage{amsmath,amssymb}
\usepackage{graphicx}
\usepackage{float}
\usepackage{placeins}
\usepackage{booktabs}
\usepackage{multirow}
\usepackage{url}
\usepackage{xcolor}
\usepackage{hyperref}
\hypersetup{colorlinks=true, linkcolor=blue!60!black, citecolor=blue!60!black, urlcolor=blue!60!black}
\usepackage{doi}
\providecommand{\keywords}[1]{\par\medskip\noindent\textbf{\textit{Keywords:}} #1}

\renewcommand{\shorttitle}{Quantum Error Management in Practice}
\renewcommand{\headeright}{Preprint}
\renewcommand{\undertitle}{Preprint}

\title{Quantum Error Management in Practice: A Cross-Stack Benchmark}

\author{%
  Daniel Sierra-Sosa\thanks{Corresponding author. sierrasosa@cua.edu} \\
  Department of Computer Science\\
  The Catholic University of America\\
  Washington, DC, USA\\
  \And
  Begonya Garcia-Zapirain \\
  eVIDA Research Group\\
  University of Deusto\\
  Bilbao, Spain\\
  \And
  Cristian Marquez \\
  Department of Systems and Computing Engineering\\
  Universidad de los Andes\\
  Bogota, Colombia\\
  \And
  Kelly Garces \\
  Department of Systems and Computing Engineering\\
  Universidad de los Andes\\
  Bogota, Colombia\\
}

\begin{document}
\maketitle

\begin{abstract}
Quantum processors have crossed the one-hundred-qubit mark, but every circuit they run is still degraded by noise, and full quantum error correction remains too expensive for routine use. In the meantime, a growing commercial ecosystem promises to extract more value from today's hardware through error suppression, which prevents errors during execution, and error mitigation, which removes their biases in post-processing. IBM ships both capabilities inside its Qiskit Runtime primitives, while specialist companies such as Q-CTRL and Qedma sell managed pipelines through the Qiskit Functions Catalog. Independent comparisons of these offerings on identical workloads and identical hardware are scarce, which leaves practitioners guessing about what each layer of the stack is worth. Here we present such a comparison, executed on \texttt{ibm\_pittsburgh}, a 156-qubit IBM Quantum Heron r3 processor. On the Sampler side we run Bernstein--Vazirani, quantum phase estimation, GHZ-state preparation, and randomized mirror circuits at up to 100 measured qubits, comparing raw execution, IBM measurement twirling, and Q-CTRL's Performance Management. On the Estimator side we measure chain-averaged magnetization and correlation observables of a fixed eight-layer transverse-field Ising circuit at 25, 50, and 75 qubits against an exact matrix-product-state reference, comparing raw execution, IBM's TREX-plus-twirling configuration, Q-CTRL, and Qedma's QESEM. On the three structured Sampler workloads, Q-CTRL's reported QPU times remained within the same order as those of the IBM configurations. Across the six TFIM observable--size cases, the aggregate mean absolute error was \(0.0883\) for IBM raw execution, \(0.0807\) for IBM TREX plus twirling, \(0.0285\) for Q-CTRL, and \(0.0188\) for QESEM. Relative to raw execution, Q-CTRL and QESEM reduced aggregate error by factors of \(3.10\) and \(4.70\), respectively, while QESEM used \(7.5\)--\(11.1\times\) the reported QPU time of Q-CTRL. IBM's TREX-plus-twirling configuration improves the magnetization but systematically underestimates the correlator. The uncertainty-related fields returned by the providers are retained for transparency but are not compared because their statistical definitions are provider-specific. All results derive from a seeded, job-level, submit-and-collect workflow designed for reproducibility.
\end{abstract}


\FloatBarrier
\section{Introduction}
\label{sec:intro}

Quantum computers have reached a curious stage in their development. Superconducting processors with more than one hundred qubits are available through the cloud, and circuits acting on all of those qubits can be compiled and executed in seconds. Yet, the results of such circuits are never what an ideal quantum computer would return. Each two-qubit gate on the best current devices fails with a probability of roughly one in a thousand, qubits lose their quantum character within a few hundred microseconds, and the readout of every qubit misreports its state a percent or so of the time. Multiply these small failure probabilities across the thousands of operations in a useful circuit and the ideal signal is quickly buried. This is the regime Preskill named noisy intermediate-scale quantum, or NISQ~\cite{preskill2018}, and the central question of the field is how much useful computation can be extracted from it. The question is not academic: application communities already prototype real workloads on these processors, from combinatorial optimization~\cite{giraldo2026opt} to quantum machine learning~\cite{sierrasosa2020tfq,maheshwari2022vqc}, and each of these results inherits the noise of the execution layer beneath it.

The principled long-term answer is quantum error correction, in which logical quantum information is encoded redundantly across many physical qubits so that errors can be detected and reversed faster than they accumulate~\cite{shor1995,steane1996,fowler2012}. Experimental progress here has been remarkable: logical qubits whose error rate improves as the code grows~\cite{google2023}, a below-threshold surface-code memory~\cite{google2025}, and high-rate codes that promise the same protection with roughly ten times fewer physical qubits~\cite{bravyi2024}. Even so, every one of these demonstrations protects memories or very small logical circuits. Running an algorithm of practical size under error correction still requires physical-qubit counts and control overheads that lie years beyond present devices. Whoever wants a correct answer from a quantum computer in this decade must therefore fight noise by other means.

Two complementary families of techniques fill that gap, and the distinction between them organizes everything in this paper. Error \emph{suppression} acts before and during execution to prevent errors from happening in the first place. Its tools include noise-aware qubit placement and circuit compilation, dynamical decoupling sequences that echo away slow environmental drift on idling qubits~\cite{viola1999}, and randomized compiling or Pauli twirling, which converts structured coherent errors into effectively random ones that accumulate more benignly~\cite{wallman2016}. Suppression adds little or no sampling overhead, and because it improves the physical execution itself, it helps every kind of output, including the raw bitstrings a sampling algorithm produces. Error \emph{mitigation}, by contrast, accepts that the execution is noisy and removes the resulting bias afterwards, in classical post-processing. Zero-noise extrapolation runs the circuit at deliberately amplified noise levels and extrapolates the measured expectation values back to the zero-noise limit~\cite{temme2017,li2017,kandala2019}. Probabilistic error cancellation learns a model of the device noise and inverts it as a quasi-probabilistic sum of circuits, yielding unbiased estimates at an exponential sampling cost~\cite{vandenberg2023}. Twirled readout error extinction, known as TREX, symmetrizes measurement errors and removes their bias from expectation values with a small calibration overhead~\cite{vandenberg2022}. Mitigation techniques of this kind were essential to IBM's demonstration that a 127-qubit processor can compute accurate expectation values beyond brute-force classical simulation~\cite{kim2023}, and they are reviewed comprehensively in Ref.~\cite{cai2023}. The crucial limitation is that mitigation reconstructs \emph{expectation values} of observables; it cannot repair individual output bitstrings. The crucial limitation is that error mitigation cannot retrospectively
identify the error-free value of an individual observed bitstring, although estimated distributions over bitstrings can be mitigated in post-processing. In this study, the Sampler track evaluates provider-returned counts without applying a separate distribution-level mitigation procedure, whereas the Estimator track evaluates both suppression and mitigation of observable expectation values.

These techniques reach users through an increasingly commercial software stack, and the present work benchmarks its three most visible providers. The foundation is IBM Quantum itself. Its Qiskit Runtime service exposes two execution primitives, a Sampler that returns measured bitstrings and an Estimator that returns expectation values, and both accept error-handling options that the user can switch on individually or through aggregate resilience levels~\cite{qiskit2024,ibmoptions}. Dynamical decoupling, Pauli twirling of gates, and twirling of measurements are available in both primitives; the Estimator additionally offers TREX readout mitigation at resilience level~1 and zero-noise extrapolation at level~2, with probabilistic error cancellation as a further option, each raising accuracy at a growing cost in QPU time~\cite{ibmoptions}. In 2024 IBM added a second layer of abstraction, the Qiskit Functions Catalog, through which partner companies operate complete managed pipelines as cloud services: the user submits an abstract, untranspiled circuit together with a target, and the function returns corrected results, handling qubit selection, compilation, suppression, mitigation, and execution internally~\cite{ibmfunctions}.

Two such partners are evaluated here. Q-CTRL, a quantum infrastructure software company headquartered in Sydney, offers its Fire Opal technology on IBM hardware as the Performance Management function~\cite{ibmqctrl}. The pipeline is fully autonomous: it performs error-aware layout and compilation, optimizes gate implementations, applies dynamical decoupling and measurement-error suppression, and returns either bitstring distributions or expectation values without any user configuration. Because these steps are almost entirely suppression, they add essentially no sampling overhead, and published benchmarks report success-probability improvements of up to three orders of magnitude on algorithmic circuits~\cite{mundada2023}. Qedma, founded in Tel Aviv by Asif Sinay, Dorit Aharonov, and Netanel Lindner, takes the complementary, mitigation-centric route with its Quantum Error Suppression and Error Mitigation software, QESEM, likewise available as a Qiskit Function~\cite{ibmqesem,qedma2025em}. QESEM first characterizes the noise of the specific gate layers appearing in the submitted circuit, suppresses part of that noise during execution, and removes the remainder through an unbiased mitigation step in post-processing; the user specifies a target statistical precision and a QPU-time budget, and the function returns expectation values accompanied by calibrated uncertainty estimates~\cite{aharonov2025qesem}. QESEM operates exclusively through the Estimator interface, since unbiased mitigation is defined for observables rather than for bitstrings.

Each of these vendors publishes performance figures, but almost always on workloads of their own choosing, on different devices, and at different times, which makes the numbers difficult to compare and impossible to convert into practical guidance. Independent protocol-level benchmarks of individual processors do exist, including a teleportation-based variant we introduced to quantify the execution reliability of a single QPU~\cite{marquez2025tp}, but they characterize the bare device rather than the commercial software stacks layered on top of it. What a practitioner actually needs to know is: for a
given class of circuits at a given scale, how much accuracy does each stack deliver, and at what cost in quantum-processor time? Answering that question requires an independent benchmark that runs the same logical circuits, on the same processor, within the same time window, under a protocol that treats every provider the way its vendor intends it to be used.

This paper provides such a benchmark. All experiments were run on \texttt{ibm\_pittsburgh}, a 156-qubit Heron r3 processor. IBM and Q-CTRL instances used matched requested budgets of \(2^{15}=32{,}768\) shots per benchmark instance, whereas QESEM was
operated through its provider-native interface with a target precision of \(0.1\) and a maximum QPU-time budget of \(600\,\mathrm{s}\) per job. A seeded submit-and-collect workflow records the logical circuits, provider configurations, and job identifiers used in the study. On the Sampler side we compare raw execution, IBM measurement twirling, and Q-CTRL on four algorithm families with unambiguous ideal outputs: Bernstein--Vazirani at 25 to 75 qubits, quantum phase estimation at 10 to 30 counting qubits, GHZ-state preparation at 25 and 50 qubits, and randomized mirror circuits at 25 to 100 qubits. On the Estimator side we compare raw execution, IBM's TREX-plus-twirling configuration, Q-CTRL, and QESEM on chain-averaged magnetization and correlation observables of a fixed eight-layer transverse-field Ising circuit at 25, 50, and 75 qubits, scored against a numerically exact matrix-product-state reference. We also retain the \texttt{stds} fields returned by each provider for transparency, but we do not use them in cross-provider comparisons because Qiskit Estimator implementations do not enforce a common statistical
definition for this field.

The remainder of the paper is organized as follows. Section~\ref{sec:methods} describes the hardware, the common execution protocol, the physics and construction of every benchmark circuit, the scoring metrics, and the exact configuration of each provider. Section~\ref{sec:results} presents the Sampler and Estimator results in turn and closes with a synthesis of the cost-versus-accuracy trade-off. Section~\ref{sec:conclusions} summarizes the findings, states their limitations, and offers guidance for practitioners choosing among these stacks today.

\FloatBarrier
\section{Materials and methods}
\label{sec:methods}

\FloatBarrier
\subsection{Hardware, primitives, and common execution protocol}
\label{sec:common}

All circuits in this study were executed on \texttt{ibm\_pittsburgh}, a 156-qubit IBM Quantum processor of the Heron r3 generation~\cite{ibmpittsburgh}. Heron devices arrange fixed-frequency transmon qubits on a heavy-hexagonal lattice and couple neighboring qubits through tunable couplers, which suppresses parasitic crosstalk and makes the controlled-Z gate the native two-qubit operation. The generation is characterized by coherence times in the hundreds of microseconds and median two-qubit error rates at the level of a few parts in a thousand, which places circuits of a few thousand entangling gates at the edge of what survives unaided and therefore squarely in the regime where suppression and mitigation matter.

Every benchmark instance was executed using the version-2 Qiskit Runtime primitives. IBM and Q-CTRL instances used a matched requested budget of \(2^{15}=32{,}768\) shots per benchmark instance. QESEM instances were instead configured through its provider-native precision-controlled interface, using a target precision of \(0.1\) and a maximum QPU-time budget of \(600\,\mathrm{s}\) per job. Consequently, IBM and Q-CTRL are compared under matched requested-shot budgets, whereas QESEM represents a precision- and QPU-time-constrained operating point. For the Sampler results, \(32{,}768\) shots bound the worst-case binomial standard error of a reported success probability to approximately \(0.28\) percentage points. This makes finite-shot fluctuations small relative to the largest observed success-probability differences, but it does not eliminate calibration drift, compilation differences, or other job-to-job hardware variability.

The providers differ in where they take control of the circuit, and the execution protocol respects that difference deliberately. IBM's own primitives accept only circuits already expressed in the instruction set architecture of the target device, so for the raw and twirled IBM configurations we transpiled every abstract circuit with Qiskit's preset pass manager at its highest optimization level, level~3, which performs layout, routing, and gate synthesis for \texttt{ibm\_pittsburgh}. The managed functions of Q-CTRL and QESEM, by contrast, are products whose value proposition explicitly includes qubit selection and compilation, so they received the abstract, untranspiled circuits and were left to make their own choices. The comparison is therefore like-for-like at the service boundary: every provider sees the same logical workload and is exercised exactly as its vendor intends.

Because queue times on a shared 156-qubit machine are unpredictable, the workflow separates submission from analysis. For each track, all jobs across all providers were submitted in a single session, and every job identifier, together with the seed and configuration that produced it, was recorded in a JSON manifest; results were collected and scored later, offline, from those identifiers. A master seed of 42 fixes the Bernstein--Vazirani hidden strings and all other pseudo-random choices, and the randomized mirror circuits draw per-instance seeds from a base value of 4242, so that every circuit in the study can be regenerated bit-for-bit from the manifest alone. This submit-and-collect design keeps the comparison and reproducible, and confines calibration drift to the gap between provider submissions rather than letting it leak into the analysis.

\FloatBarrier
\subsection{Sampler benchmark}
\label{sec:sampler}

\FloatBarrier
\subsubsection{Design of the workload}

The Sampler primitive returns the raw currency of quantum computation, measured bitstrings, and a meaningful Sampler benchmark therefore needs circuits whose ideal output distribution is known exactly and is concentrated enough that success can be read off without ambiguity. We chose four algorithm families that satisfy this requirement while probing complementary failure modes of the hardware. Two of them, Bernstein--Vazirani and quantum phase estimation, are structured textbook algorithms whose ideal output is a single bitstring, the first shallow and the second deep. The third, GHZ-state preparation, produces maximal multipartite entanglement and ideally only two bitstrings. The fourth, randomized mirror circuits, is an unstructured stress test built from the native gates of the device. Table~\ref{tab:workloads} summarizes the measured widths, the physical register sizes, and the scoring rule for each family, and the following subsections describe each circuit in turn, together with its benchmark parameters and an illustrative instance.

\begin{table}[!htb]
\centering
\caption{Sampler workloads. Every instance was executed with $2^{15}$ shots by every provider configuration. The score modes refer to the metrics defined in Section~\ref{sec:sampmetrics}.}
\label{tab:workloads}
\begin{tabular}{lllll}
\toprule
Algorithm & Measured widths $n$ & Physical width & Ideal output & Score mode \\
\midrule
BV  & 25, 50, 75      & $n+1$ & seeded hidden string $s$        & exact, Eq.~\eqref{eq:exact} \\
QPE & 10, 20, 30      & $n+1$ & binary phase $0100\cdots0$ ($\theta=\tfrac14$) & exact, Eq.~\eqref{eq:exact} \\
RND & 25, 50, 75, 100 & $n$   & balanced nonzero initial string & exact, Eq.~\eqref{eq:exact} \\
GHZ & 25, 50          & $n$   & $0\cdots0$ or $1\cdots1$        & valid set, Eq.~\eqref{eq:validset} \\
\bottomrule
\end{tabular}
\end{table}

\FloatBarrier
\subsubsection{Bernstein--Vazirani}

The Bernstein--Vazirani algorithm~\cite{bernstein1997} is the simplest quantum routine with a provable query advantage, and it opens the suite because its physics is transparent from end to end. An oracle hides an $n$-bit string $s$ behind the Boolean function $f(x) = s \cdot x \bmod 2$: classically each query reveals at most one bit of information about $s$, so $n$ queries are unavoidable, whereas the quantum algorithm recovers the entire string from a single one. The mechanism is phase kickback. A layer of Hadamard gates prepares the query register in the uniform superposition over all inputs while an ancilla is prepared in the state $|-\rangle$; a single oracle call then leaves the register in $\tfrac{1}{\sqrt{2^n}}\sum_x (-1)^{s\cdot x}\,|x\rangle$, with the answer written into a pattern of signs rather than into any measurable population. Because the oracle function is linear, that sign pattern is exactly the Hadamard transform of the basis state $|s\rangle$, so a closing Hadamard layer converts it back and the register reads $s$ with probability one. In circuit form the oracle is a cascade of CNOT gates, one from each qubit at which $s_i = 1$ onto the shared ancilla.

\begin{figure}[!htbp]
\centering
\includegraphics[width=0.8\textwidth]{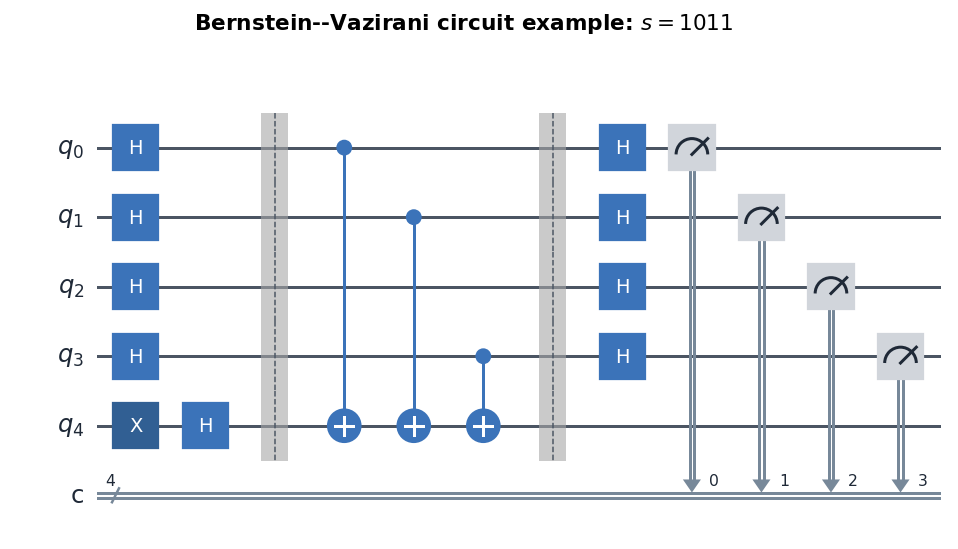}
\caption{Illustrative Bernstein--Vazirani circuit as generated by the benchmark code. Hadamard layers sandwich an oracle made of CNOT gates onto a phase-kickback ancilla, and the ideal measurement returns the hidden string $s$ with probability one.}
\label{fig:bvcirc}
\end{figure}

Two hardware-facing features make the family a sharp benchmark despite its textbook simplicity. First, although the logical circuit is only three single-qubit layers around one oracle, every oracle CNOT shares the ancilla as its target, so the entangling gates cannot act simultaneously and the executed depth grows with the Hamming weight of $s$, roughly $n/2$ under our seeding. Second, on a heavy-hexagonal lattice each qubit has at most three neighbors, so the transpiler must shuttle distant controls toward the ancilla, inflating the executed two-qubit count well beyond its logical value. Wide Bernstein--Vazirani instances therefore measure routing quality, two-qubit fidelity, and full-register readout together, concentrated on a single deterministic outcome. The benchmark parameters are the following: measured widths $n = 25$, $50$, and $75$, each instance occupying $n+1$ physical qubits including the ancilla; hidden strings drawn reproducibly from the master seed 42, with expected Hamming weight $n/2$, so that the oracle contains on average $n/2$ CNOT gates; and the common budget of $2^{15}$ shots. Figure~\ref{fig:bvcirc} shows a small instance of the resulting circuit.

\FloatBarrier
\subsubsection{Quantum phase estimation}

Quantum phase estimation, introduced by Kitaev~\cite{kitaev1995} and the engine behind most exponential quantum speedups~\cite{nielsen2010}, plays the opposite role in the suite: it is deep where Bernstein--Vazirani is shallow, and it probes precision where the former probes breadth. Given a unitary $U$ with eigenstate $|u\rangle$ and eigenphase $\theta$, so that $U|u\rangle = e^{2\pi i\theta}|u\rangle$, the algorithm prepares $m$ counting qubits in uniform superposition and applies the controlled powers $U^{2^{k}}$ for $k = 0, \dots, m-1$, after which the counting register holds the phase gradient $\tfrac{1}{\sqrt{2^m}}\sum_{k=0}^{2^m-1} e^{2\pi i \theta k}\,|k\rangle$; an inverse quantum Fourier transform then converts that gradient into the computational-basis value $2^m\theta$. We instantiate $U$ as the single-qubit phase gate $P(2\pi\theta)$ with $\theta = 1/4$ acting on its eigenstate $|1\rangle$. Because $1/4$ has an exact two-bit binary expansion, the ideal outcome is the single string $0100\cdots0$ with certainty at every register size, which keeps the exact-output metric meaningful, and because every controlled-phase angle is computed as an exact rational multiple of $2\pi$ and reduced modulo $2\pi$, no floating-point drift accumulates. For this eigenphase, in fact, every controlled power beyond $U^{2}$ reduces to the identity, so the phase-encoding stage of the circuit is almost free.

\begin{figure}[!htbp]
\centering
\includegraphics[width=0.95\textwidth]{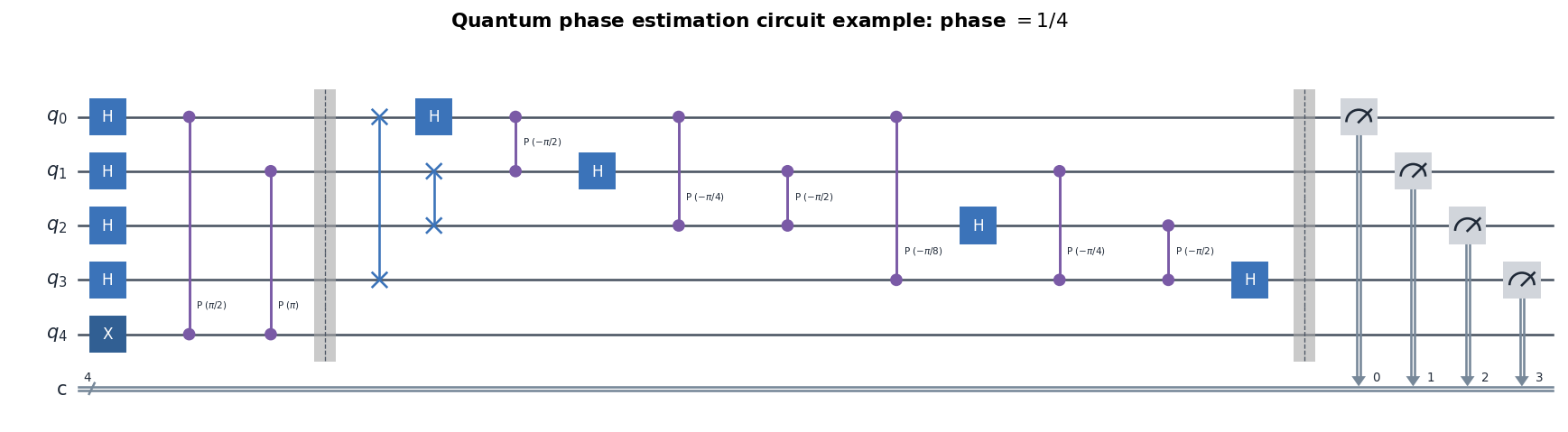}
\caption{Quantum-phase-estimation circuit for $m = 4$ counting qubits and phase $\theta = 1/4$. Controlled powers of the phase unitary are followed by an explicit inverse quantum Fourier transform, and the ideal outcome is the single bitstring encoding the binary expansion of $\theta$.}
\label{fig:qpecirc}
\end{figure}

The difficulty is concentrated instead in the explicit, gate-level inverse Fourier transform, which contains $m(m-1)/2$ controlled-phase rotations with angles descending geometrically to $\pi/2^{\,m-1}$ and which couples every counting qubit to every other. Rotations this small are non-Clifford and are exquisitely sensitive to calibration and coherent-control errors, while the all-to-all coupling pattern forces long SWAP chains on a sparse lattice. The result is a circuit whose executed two-qubit content grows quadratically with $m$ and which stresses precisely what a shallow oracle cannot: the ability of the device to sustain a long, coherent sequence of precise small-angle entangling operations. The benchmark parameters are the following: counting registers of $m = 10$, $20$, and $30$ qubits, each instance occupying $m+1$ physical qubits including the eigenstate qubit; the fixed eigenphase $\theta = 1/4$ with ideal output $0100\cdots0$; and the common budget of $2^{15}$ shots. Figure~\ref{fig:qpecirc} shows the structure at $m = 4$.

\FloatBarrier
\subsubsection{GHZ-state preparation}

The Greenberger--Horne--Zeilinger state~\cite{ghz1989} tests something neither structured algorithm does: the creation and survival of entanglement shared coherently by the entire register. A single Hadamard places the first qubit in superposition, and a ladder of $n-1$ CNOT gates then copies that superposition down the chain, producing $(|0\cdots0\rangle + |1\cdots1\rangle)/\sqrt{2}$, a state so strongly entangled that measuring any one qubit collapses all the others. The construction is as simple as an entangling circuit can be, but its execution profile is unforgiving. The superposition propagates sequentially, so the entangling depth is $n-1$ along a hardware path; the early qubits then idle for a time that grows linearly with $n$ while the ladder completes, accumulating decoherence; and the coherence between the two macroscopic branches is degraded by every dephasing event anywhere in the register. GHZ preparation is for these reasons a standard probe of coherent scale on superconducting processors~\cite{mooney2021}, and the linear ladder embeds naturally along a path of the heavy-hexagonal lattice.

\begin{figure}[!htbp]
\centering
\includegraphics[width=0.62\textwidth]{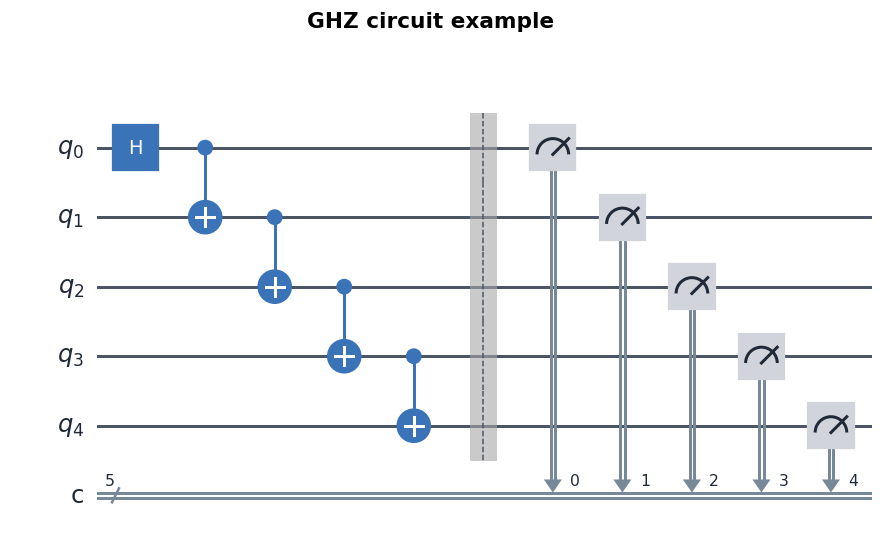}
\caption{GHZ preparation on five qubits: a Hadamard followed by a CNOT ladder. The ideal output distribution assigns probability one half to each of the all-zeros and all-ones bitstrings.}
\label{fig:ghzcirc}
\end{figure}

The reported quantity is therefore a support-population metric rather than a GHZ-state fidelity or coherence witness. It is sensitive to leakage into bitstrings outside the two ideal branches, but it can overestimate state quality under branch imbalance or relaxation toward the all-zero state. Certifying genuine multipartite coherence would require additional measurements, such as parity oscillations, which are outside the scope of this study~\cite{mooney2021}. The benchmark parameters are the following: register sizes $n = 25$ and $50$, each circuit consisting of one Hadamard and $n-1$ CNOT gates laid along a hardware path, with no ancillas and no free parameters; scoring by the valid-set probability of Eq.~\eqref{eq:validset}. Figure~\ref{fig:ghzcirc} shows the five-qubit ladder.

\FloatBarrier
\subsubsection{Randomized mirror circuits}

The randomized mirror circuits complete the suite with an unstructured stress test in the spirit of mirror-circuit benchmarking~\cite{proctor2022}, whose central idea solves a genuine problem of benchmarking at scale: a random 100-qubit circuit cannot be simulated classically to learn its ideal output, but a random circuit followed by its exact inverse has the identity as its ideal unitary, so the correct outcome is known by construction at any width, an inversion strategy we have also exploited in a teleportation-based benchmark of single-QPU reliability~\cite{marquez2025tp}. Each instance here begins with a layer of $X$ gates that imprints a target bitstring, applies a four-layer body alternating uniformly random single-qubit gates from the native pool $\{X, \sqrt{X}, R_Z(\pm\pi/2)\}$ with brickwork patterns of controlled-Z entanglers on alternating even and odd bonds, and then appends the gate-by-gate inverse of that body, so that an ideal device returns the target string on every shot after eight entangling layers.

\begin{figure}[!htbp]
\centering
\includegraphics[width=\textwidth]{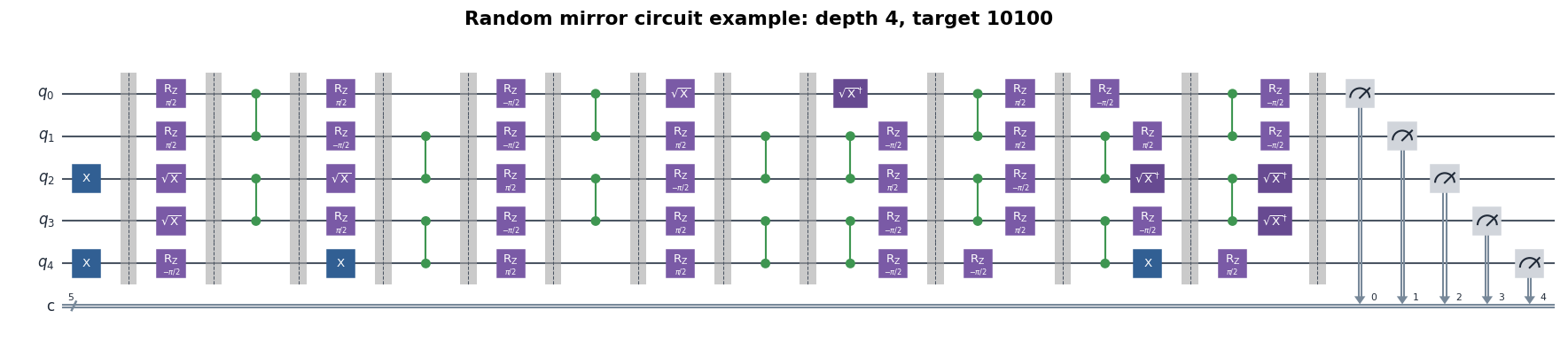}
\caption{Randomized mirror circuit on five qubits: an initial $X$ layer sets a balanced nonzero target, a four-layer body of random native single-qubit gates and controlled-Z brickwork follows, and the inverted body returns the ideal state to the target string.}
\label{fig:rndcirc}
\end{figure}

Three design choices shape what the family measures. Because the logical circuit is expressed in the processor's native gate
family, IBM-side transpilation requires less gate synthesis than for the other workloads. Layout selection, routing, scheduling, and circuit simplification nevertheless remain possible, so the family probes end-to-end execution of dense native-gate layers rather than isolating bare gate fidelity in complete independence of compilation. Because the target string is balanced and nonzero, relaxation toward the all-zero state is less likely to receive spurious credit than it would for an all-zero target. This construction reduces, but does not eliminate, the directional effect of \(T_1\) relaxation. And because the inverse is the exact reversal of the body with no randomizing layer in between, certain coherent errors can partially cancel between a gate and its inverse, a known property of mirror constructions~\cite{proctor2022}, so mirror success is best read as an optimistic but well-defined proxy for forward-circuit fidelity. The benchmark parameters are the following: widths $n = 25$, $50$, $75$, and $100$, the largest register in the study; per-instance gate choices and target strings drawn reproducibly from seeds derived from the base value 4242; and the common budget of $2^{15}$ shots. Figure~\ref{fig:rndcirc} shows a five-qubit example.

\FloatBarrier
\subsubsection{Success metrics}
\label{sec:sampmetrics}

For the three families whose ideal output is a single bitstring $s^{*}$, success is simply the empirical probability of observing it,
\begin{equation}
P_{\mathrm{exact}} \;=\; \frac{C(s^{*})}{N_{\mathrm{shots}}},
\label{eq:exact}
\end{equation}
where $C(s^{*})$ counts the shots that returned $s^{*}$ and $N_{\mathrm{shots}} = 2^{15}$. This is the most demanding sampling metric available, because a single flipped bit anywhere in the register disqualifies a shot, and it is also the operationally relevant one: It measures the fraction of shots that exactly match the target bitstring. This is stricter than asking whether the target is merely the modal output of the histogram. For the GHZ family the ideal distribution is supported on two strings, and the natural generalization scores the mass landing on that support,
\begin{equation}
P_{\mathrm{valid}} \;=\; \frac{C(0^{n}) + C(1^{n})}{N_{\mathrm{shots}}},
\label{eq:validset}
\end{equation}
which reaches one for a perfect device and $2^{1-n}$ for uniform noise. With the fixed shot budget, both metrics resolve probabilities down to $1/2^{15} \approx 0.003\%$, a floor that becomes relevant when raw phase estimation collapses in Section~\ref{sec:sampres}.

In addition to the raw success probability, we report the effective per-qubit success rate $P^{1/n}$ in Section~\ref{sec:sampres}. This normalization expresses the global success probability on a common per-qubit scale, making degradation across circuit sizes easier to compare. It is an aggregate figure of merit and should not be interpreted as the measured contribution of an individual physical qubit or as evidence that qubit errors are independent. The uniform-noise baselines $2^{-n}$ or $2^{1-n}$ for the valid set of GHZ are negligible at the widths studied here, so the exact probabilities of the raw output are not significantly inflated by chance.

\FloatBarrier
\subsubsection{Provider configurations}
\label{sec:sampcfg}

Three configurations received the identical workload, and Table~\ref{tab:sampcfg} summarizes them. The first, which we call IBM raw, is the control arm: the level-3 transpiled circuit runs through SamplerV2 with every optional error-handling feature switched off, including dynamical decoupling and all twirling, so that it exposes the unassisted physics of the device. The second, IBM measurement twirling, enables exactly one option on top of the control: random $X$ flips are inserted immediately before measurement and undone in classical post-processing, which symmetrizes the readout error across the register~\cite{ibmoptions}. Gate twirling was disabled to isolate the effect of measurement twirling under a consistent configuration across all four circuit families. The third configuration submits the abstract circuits to Q-CTRL's Performance Management function with its default, fully autonomous pipeline, which internally performs error-aware layout, optimized compilation, dynamical decoupling, and measurement-error suppression before returning bitstring counts~\cite{ibmqctrl,mundada2023}. Qedma's QESEM does not appear in this track because it exposes no Sampler interface: as an unbiased mitigation method it is defined for expectation values, a boundary discussed in Section~\ref{sec:intro} and one that this benchmark respects rather than works around.

\begin{table}[!htb]
\centering
\caption{Provider configurations for the Sampler track. Every configuration executed the same logical circuits with $2^{15}$ shots.}
\label{tab:sampcfg}
\begin{tabular}{lp{4.1cm}p{6.3cm}}
\toprule
Configuration & Interface & Error-management settings \\
\midrule
IBM raw & Runtime \texttt{SamplerV2}, ISA circuits (O3) & gate twirling off; measurement twirling off; dynamical decoupling off \\
\midrule
IBM measurement twirling & Runtime \texttt{SamplerV2}, ISA circuits (O3) & measurement twirling on; gate twirling off; dynamical decoupling off \\
\midrule
Q-CTRL & Qiskit Function \texttt{q-ctrl/}\allowbreak\texttt{performance-management}, abstract circuits & full automated suppression pipeline (layout, compilation, dynamical decoupling, measurement-error suppression and processing) \\
\bottomrule
\end{tabular}
\end{table}

\FloatBarrier
\subsection{Estimator benchmark}
\label{sec:estimator}

\FloatBarrier
\subsubsection{The transverse-field Ising workload: circuit, observables, and exact reference}
\label{sec:tfim}

The Estimator track asks a different question from the Sampler track: given the same circuit and the same observables, which stack returns expectation values closest to the exact answer, and at what quantum-processor cost? Answering it requires a circuit that is classically verifiable at every size we run, physically meaningful rather than contrived, and genuinely sensitive to hardware noise. The one-dimensional transverse-field Ising model meets all three requirements and has become the de facto standard for utility-scale estimation experiments~\cite{kim2023,aharonov2025qesem}.

The model describes a chain of $n$ spins with the Hamiltonian
\begin{equation}
H \;=\; -J \sum_{i=1}^{n-1} Z_i Z_{i+1} \;-\; h \sum_{i=1}^{n} X_i,
\label{eq:tfim}
\end{equation}
in which the first term favors ferromagnetic alignment of neighboring spins along $Z$ and the second is a transverse magnetic field that makes the aligned configurations quantum-mechanically unstable. The competition between the two terms produces the textbook example of a quantum phase transition, exactly solvable in one dimension~\cite{pfeuty1970} and the opening chapter of the modern theory of quantum criticality~\cite{sachdev2011}. Simulating its dynamics on a gate-based computer proceeds by Trotterization: the evolution $e^{-iHt}$ is sliced into repeated layers, each of which applies the Ising couplings and the transverse field in alternation. One layer acts as
\begin{equation}
U_{\mathrm{layer}}=\left[\prod_{i=1}^{n} R_{X_i}(\theta_X)\right]
\left[\prod_{i\ \mathrm{odd}}
R_{Z_iZ_{i+1}}(\theta_{ZZ})\right]
\left[\prod_{i\ \mathrm{even}}R_{Z_iZ_{i+1}}(\theta_{ZZ})
\right],
\label{eq:trotter}
\end{equation}
where $R_{ZZ}(\theta) = e^{-i\theta\, Z\otimes Z/2}$ and $R_X(\theta) = e^{-i\theta\, X/2}$, and the even and odd bonds are applied in two sublayers so that all two-qubit gates within a sublayer act on disjoint pairs. The rotation angles map onto the Hamiltonian couplings through the Trotter step, $\theta_{ZZ} = -2J\,\delta t$ and $\theta_{X} = -2h\,\delta t$, so a fixed-angle circuit of this form is exactly the kicked Ising dynamics used in IBM's utility experiment~\cite{kim2023}.

\begin{figure}[!htbp]
\centering
\includegraphics[width=0.8\textwidth]{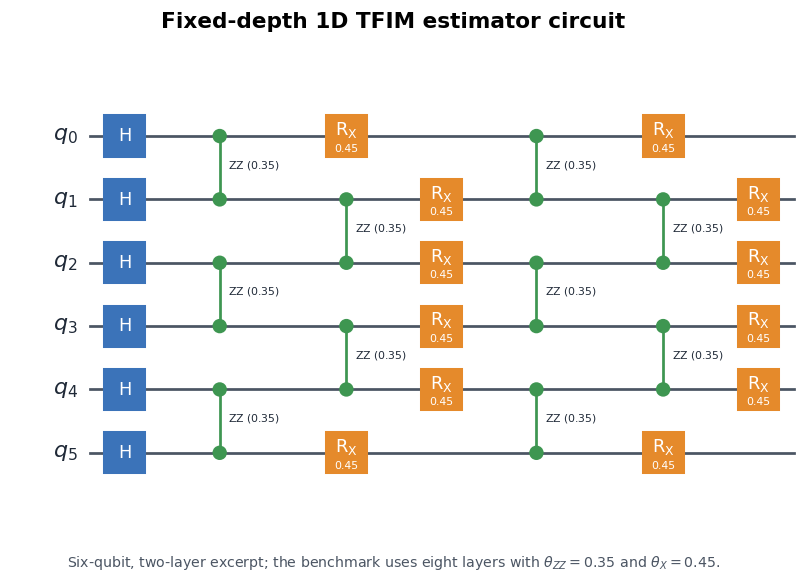}
\caption{Excerpt of the transverse-field Ising benchmark circuit, showing two of the eight layers on six qubits. Each layer applies $R_{ZZ}(0.35)$ on even bonds, then on odd bonds, followed by $R_X(0.45)$ on every qubit, acting on the initial state $|+\rangle^{\otimes n}$.}
\label{fig:tfimcirc}
\end{figure}

Our benchmark instance fixes the angles at $\theta_{ZZ} = 0.35$ and $\theta_{X} = 0.45$, prepares the chain in the uniform superposition $|+\rangle^{\otimes n}$, and applies eight layers of Eq.~\eqref{eq:trotter} on open chains of $n = 25$, $50$, and $75$ qubits, each embedded along a path of the heavy-hexagonal lattice so that every $R_{ZZ}$ acts between physical neighbors. The logical interaction graph is an open chain. For the IBM Runtime configurations, the ISA circuit and mapped observables were placed on a connected physical path of the heavy-hex lattice. Q-CTRL and QESEM received the abstract circuit and selected their own physical layouts and compilation strategies.. Eight layers is deep enough that noise visibly degrades the observables, as Section~\ref{sec:estres} shows, yet shallow enough that the state remains within reach of an exact classical reference. Figure~\ref{fig:tfimcirc} shows a two-layer, six-qubit excerpt of the repeating structure.

Two chain-averaged observables are measured in the final state. The first is the transverse magnetization per site,
\begin{equation}
m_X \;=\; \frac{1}{n}\sum_{i=1}^{n} \langle X_i \rangle,
\qquad
c_{ZZ} \;=\; \frac{1}{n-1}\sum_{i=1}^{n-1} \langle Z_i Z_{i+1} \rangle,
\label{eq:obs}
\end{equation}
and the second, defined alongside it, is the average nearest-neighbor correlator. Averaging over the chain keeps the observables intensive, so their ideal values remain of order one at every system size and results at $n = 25$, $50$, and $75$ can be compared on a single axis. The pair is chosen for its complementary physics. The initial state $|+\rangle^{\otimes n}$ has maximal transverse magnetization, $m_X = 1$, and no correlations at all, $c_{ZZ} = 0$; the Ising layers then partially disorder $m_X$ while building $c_{ZZ}$ up from nothing. Many common noise channels attenuate these observables toward zero, although the magnitude and even the direction of the bias can depend on the physical noise channel. The two observables therefore provide complementary probes of the preservation of transverse polarization and the generation of nearest-neighbor correlations.

A natural third candidate, the longitudinal magnetization $\langle Z_i \rangle$, is deliberately excluded because symmetry forces it to vanish identically. The global spin flip $\mathcal{P} = X^{\otimes n}$ commutes with every gate in the circuit, since $R_X$ rotations trivially commute with $X$ and conjugating $Z \otimes Z$ by $X \otimes X$ leaves it unchanged, and the initial state $|+\rangle^{\otimes n}$ is a $+1$ eigenstate of $\mathcal{P}$. The final state therefore satisfies $\langle Z_i \rangle = \langle \mathcal{P} Z_i \mathcal{P} \rangle = -\langle Z_i \rangle = 0$ exactly, at every depth and every size. Measuring an observable whose ideal value is pinned to zero by symmetry would reward providers for returning noise-damped answers and would test nothing about accuracy, so it is omitted.

Every reported error in the Estimator track is an error with respect to a numerically exact simulation, not to another quantum run. The reference values were computed with the matrix-product-state backend of the Qiskit Aer simulator with the singular-value truncation threshold set to $10^{-16}$, which is machine precision. Matrix product states represent one-dimensional quantum states through a controllable bond dimension that bounds their entanglement content~\cite{vidal2003,schollwock2011}; because our circuit is a strictly one-dimensional, eight-layer brickwork, the entanglement across any cut remains modest, the required bond dimension stays small, and the simulation is exact for all practical purposes at all three sizes. The ideal values that anchor the benchmark, such as $m_X = 0.6606$ and $c_{ZZ} = 0.3887$ at $n = 25$, are listed alongside every measurement in Table~\ref{tab:estres}.

\FloatBarrier
\subsubsection{Accuracy metrics}
\label{sec:estmetrics}

The primary figure of merit is the absolute error of each reported expectation value against the reference, $\Delta = |\langle O \rangle_{\mathrm{reported}} - \langle O \rangle_{\mathrm{MPS}}|$. To place errors on different observables on a common, interpretable scale we also quote a normalized success score,
\begin{equation}
S_O \;=\; 100 \times \Big(1 - \frac{\Delta}{2}\Big),
\label{eq:score}
\end{equation}
which uses the fact that both chain-averaged observables have spectrum contained in $[-1, 1]$, so the largest possible error is $2$ and $S_O$ runs from $100$ for a perfect answer to $0$ for the worst one. A combined score $S_H = (S_X + S_{ZZ})/2$ summarizes each configuration in a single number. In addition, we retain the \texttt{stds} field returned by each provider and reproduce it after the $\pm$ sign in Table~\ref{tab:estres}. These values are provider-specific outputs: Qiskit Estimator implementations do not enforce a common statistical definition for \texttt{stds}. They are therefore reported for transparency only and are not compared across providers or used in the calculation of $\Delta$, $S_O$, $S_H$, or any accuracy ranking.

To summarize the case-level errors across the complete Estimator benchmark, we additionally report the mean absolute error for each provider configuration \(p\). The observable-specific and overall mean absolute errors are defined as 

\begin{equation}
\mathrm{MAE}_{p,O}
=
\frac{1}{3}
\sum_{n\in\{25,50,75\}}
\Delta_{p,O,n},
\qquad
O\in\{m_X,c_{ZZ}\},
\label{eq:MAEP0}
\end{equation}

and
\begin{equation}
\mathrm{MAE}_{p}
=
\frac{1}{6}
\sum_{n\in\{25,50,75\}}
\left(
\Delta_{p,m_X,n}
+
\Delta_{p,c_{ZZ},n}
\right).
\label{eq:MAEP}
\end{equation}

Thus, the case-level absolute errors remain the primary measurements, while the MAE provides a single aggregate summary across the three system sizes and two observables. Because both observables are dimensionless and have spectra contained in \([-1,1]\), their absolute errors are expressed on the same scale and receive equal weight.

\FloatBarrier
\subsubsection{Provider configurations}
\label{sec:estcfg}

Four configurations measured both observables at all three sizes, and Table~\ref{tab:estcfg} summarizes them. IBM raw again serves as the control, running EstimatorV2 at resilience level~0 with all twirling and mitigation disabled. The second configuration used resilience level 1 together with explicitly enabled gate twirling: gates and measurements are Pauli-twirled, and readout bias is removed with twirled readout error extinction, in which random $X$ flips before measurement diagonalize the effective readout-error matrix and a rescaling factor learned from lightweight calibration circuits removes the resulting bias from every expectation value~\cite{vandenberg2022,ibmoptions}. We also configured IBM's zero-noise extrapolation at resilience level~2, but its default three noise factors triple the QPU time of every job, and the corresponding arm was not executed within the study's compute budget; we return to this omission in the limitations. The third configuration submits the abstract circuit and observables to Q-CTRL's Performance Management estimator, whose autonomous pipeline is the same suppression stack described in Section~\ref{sec:sampcfg} operating in expectation-value mode~\cite{ibmqctrl}. The fourth submits them to QESEM with a target precision of $0.1$ and a QPU-time cap of $600$ seconds per job; the function characterizes the noise of the executed layers, suppresses part of it in execution, applies unbiased mitigation to the remainder, and returns expectation values together with its provider-defined uncertainty field~\cite{ibmqesem,aharonov2025qesem}.

\begin{table}[!htb]
\centering
\caption{Provider configurations for the Estimator track. All configurations measured both observables of Eq.~\eqref{eq:obs} at $n = 25$, $50$, and $75$.}
\label{tab:estcfg}
\begin{tabular}{lp{4.1cm}p{6.3cm}}
\toprule
Configuration & Interface & Error-management settings \\
\midrule
IBM raw & Runtime \texttt{EstimatorV2}, ISA circuit + mapped observables & resilience level 0; measurement mitigation, ZNE, PEC off; twirling off \\
\midrule
IBM TREX + twirling & Runtime \texttt{EstimatorV2}, ISA circuit + mapped observables & resilience level 2; TREX measurement mitigation on; gate + measurement twirling on; ZNE/PEC off \\
\midrule
Q-CTRL & Qiskit Function \texttt{q-ctrl/}\allowbreak\texttt{performance-management}, abstract circuit & automated error-suppression pipeline (estimator mode) \\
\midrule
QESEM & Qiskit Function \texttt{qedma/qesem}, abstract circuit & characterization-based suppression + unbiased mitigation; precision target 0.1; max execution 600 s \\
\bottomrule
\end{tabular}
\end{table}

\FloatBarrier
\section{Results}
\label{sec:results}

\FloatBarrier
\subsection{Sampler: success probabilities and cost}
\label{sec:sampres}

\begin{figure}[!htbp]
\centering
\includegraphics[width=\textwidth]{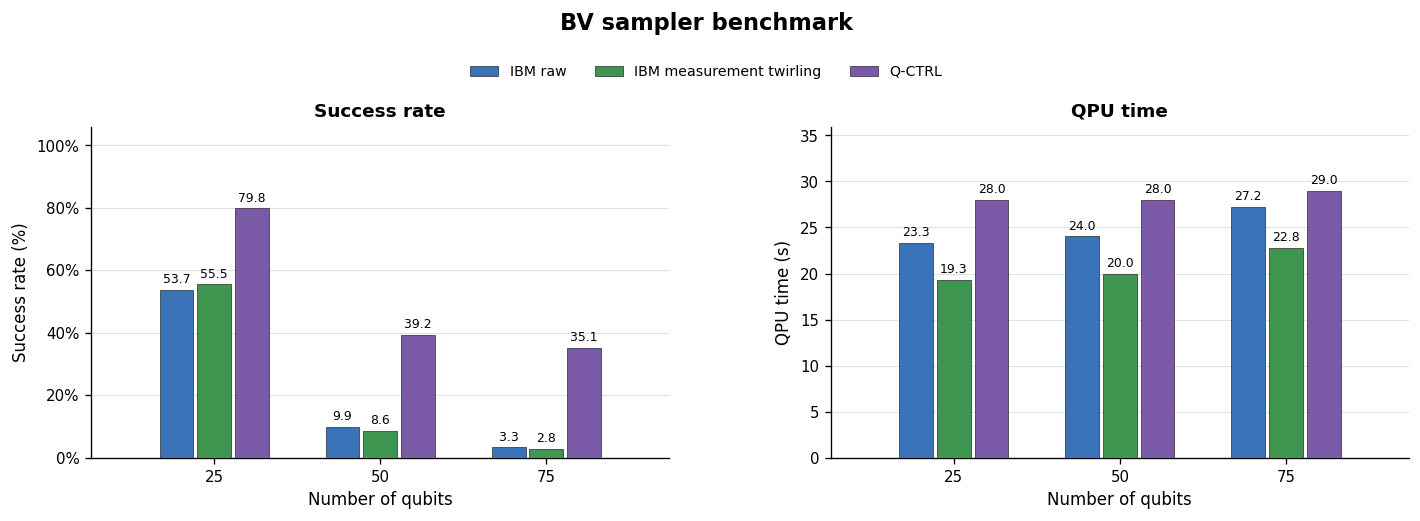}
\caption{Bernstein--Vazirani benchmark at $n = 25$, $50$, and $75$ qubits: exact-output success on the left and QPU time per job on the right.}
\label{fig:bvres}
\end{figure}

\begin{figure}[!htbp]
\centering
\includegraphics[width=\textwidth]{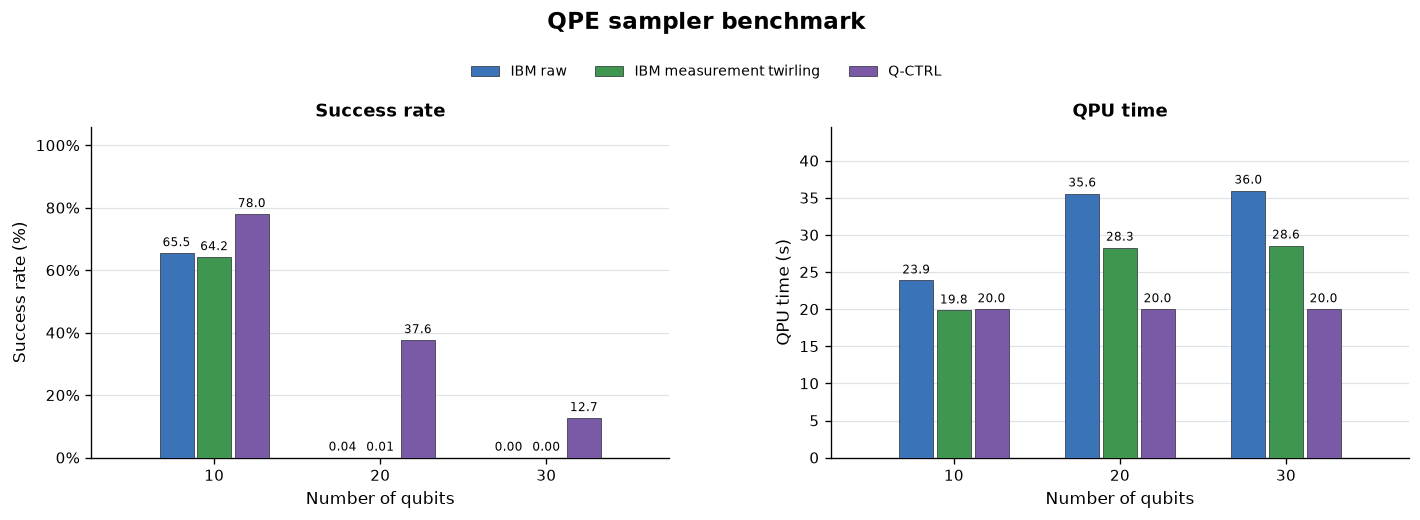}
\caption{Quantum-phase-estimation benchmark at $m = 10$, $20$, and $30$ counting qubits: exact-output success on the left and QPU time per job on the right. Raw and twirled execution are indistinguishable from zero beyond $m = 20$, while Q-CTRL retains usable success through $m = 30$.}
\label{fig:qperes}
\end{figure}

\begin{figure}[!htbp]
\centering
\includegraphics[width=\textwidth]{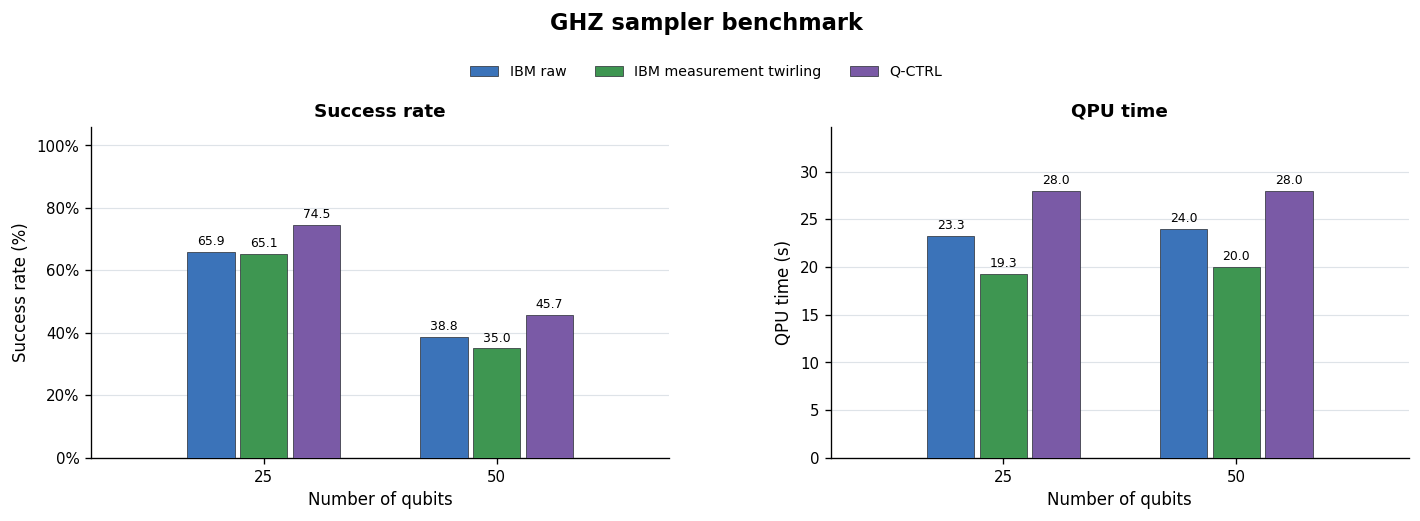}
\caption{GHZ benchmark at $n = 25$ and $50$ qubits: valid-set success probability of Eq.~\eqref{eq:validset} on the left and reported QPU time per job on the right.}
\label{fig:ghzres}
\end{figure}

\begin{figure}[!htbp]
\centering
\includegraphics[width=\textwidth]{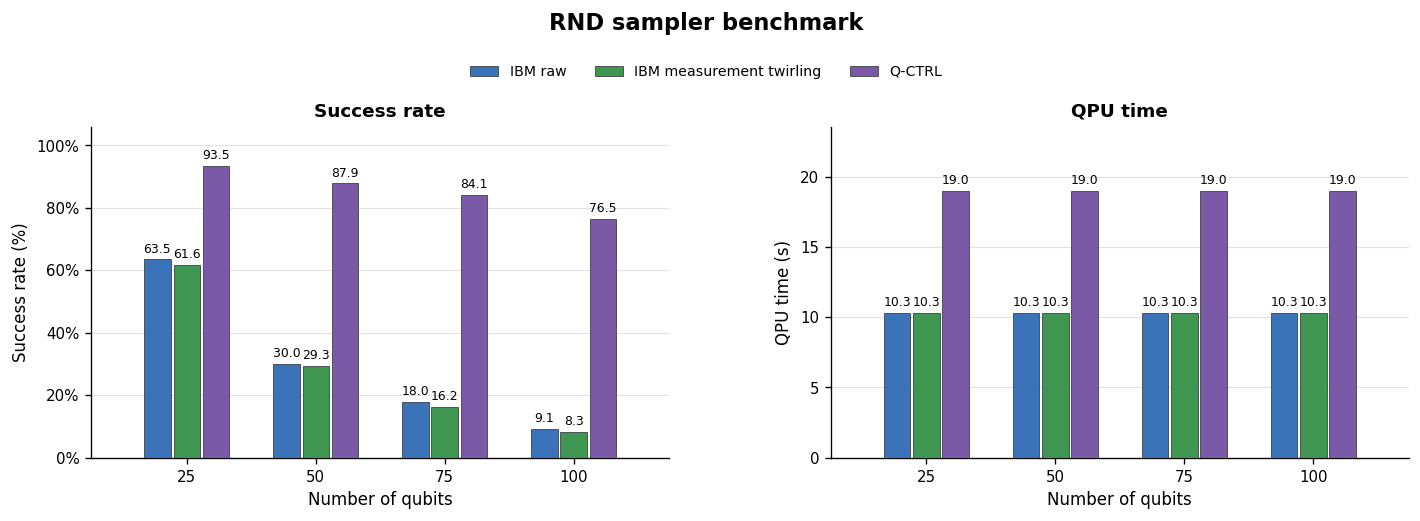}
\caption{Randomized mirror-circuit benchmark at $n = 25$, $50$, $75$, and $100$ qubits: exact-output success on the left and QPU time per job on the right.}
\label{fig:rndres}
\end{figure}

Table~\ref{tab:sampres} collects the success probabilities of all three configurations in all twelve Sampler instances, and Figures~\ref{fig:bvres} through~\ref{fig:rndres} display them family by family, each beside the paragraph that discusses it, together with the QPU time of each job. Three regularities organize the numbers: unmitigated success collapses with width at a rate set by the structure of the circuit, managed suppression rescues a large and sometimes decisive fraction of that collapse, and measurement twirling alone changes almost nothing.

\begin{table}[!htb]
\centering
\caption{Sampler success probabilities in percent, by algorithm, width, and configuration. BV, QPE, and RND report the exact-output probability of Eq.~\eqref{eq:exact}; GHZ reports the valid-set probability of Eq.~\eqref{eq:validset}. Bold marks the best configuration for each instance. The shot budget of $2^{15}$ resolves probabilities down to $0.003\%$.}
\label{tab:sampres}
\begin{tabular}{llrrr}
\toprule
Algorithm & $n$ & IBM raw & IBM twirl & Q-CTRL \\
\midrule
BV  & 25  & 53.69 & 55.47 & \textbf{79.78} \\
BV  & 50  & 9.91  & 8.56  & \textbf{39.20} \\
BV  & 75  & 3.28  & 2.83  & \textbf{35.11} \\
\midrule
QPE & 10  & 65.47 & 64.22 & \textbf{77.99} \\
QPE & 20  & 0.037 & 0.009 & \textbf{37.63} \\
QPE & 30  & 0.000 & 0.000 & \textbf{12.69} \\
\midrule
GHZ & 25  & 65.91 & 65.15 & \textbf{74.48} \\
GHZ & 50  & 38.79 & 35.00 & \textbf{45.65} \\
\midrule
RND & 25  & 63.52 & 61.61 & \textbf{93.48} \\
RND & 50  & 30.00 & 29.29 & \textbf{87.85} \\
RND & 75  & 17.96 & 16.18 & \textbf{84.12} \\
RND & 100 & 9.14  & 8.26  & \textbf{76.45} \\
\bottomrule
\end{tabular}
\end{table}

The two structured algorithms trace the steepest collapse. Raw Bernstein--Vazirani succeeds in $53.7\%$ of shots at 25 qubits, but only $3.3\%$ at 75, a decline consistent with each additional qubit contributing its share of gate and readout error to an all-or-nothing metric. Raw phase estimation is far more dramatic, because its depth grows quadratically: success falls from $65.5\%$ at ten counting qubits to $0.037\%$ at twenty, barely above the $0.003\%$ resolution floor of the shot budget, and at thirty counting qubits not a single one of the $32{,}768$ shots returned the correct string. Q-CTRL changes the picture qualitatively. It holds Bernstein--Vazirani at $79.8\%$, $39.2\%$, and $35.1\%$ across the three widths, and it keeps phase estimation alive throughout, returning $78.0\%$ at ten counting qubits, $37.6\%$ at twenty, and $12.7\%$ at thirty. At \(m=20\), the Q-CTRL success probability is approximately \(1.0\times10^{3}\) times the raw value. At \(m=30\), neither IBM configuration produced an exact success in \(32{,}768\) shots. The corresponding one-sided 95\% upper confidence bound on the underlying success probability is approximately \(0.0091\%\), whereas Q-CTRL returned \(12.69\%\). For a user who needs the correct bitstring at the top of the histogram, this is the difference between an algorithm that works and one that does not.

The GHZ support metric shows the same direction of effect at a smaller magnitude. Raw execution retains \(65.9\%\) valid-set mass at \(n=25\) and \(38.8\%\) at \(n=50\), while Q-CTRL increases these values to \(74.5\%\) and \(45.7\%\), corresponding to relative improvements of approximately \(13\%\) and \(18\%\). Because \(P_{\mathrm{valid}}\) measures only population in the two ideal computational-basis branches, these results do not determine whether Q-CTRL preserved GHZ coherence or reduced dephasing. They establish only that the managed pipeline increased the measured probability mass on the two ideal branches in these runs.

The randomized mirror circuits, the widest and least structured workload, produce the most striking absolute numbers. Raw success falls from $63.5\%$ at 25 qubits to $9.1\%$ at 100, while Q-CTRL returns $93.5\%$, $87.9\%$, $84.1\%$, and $76.5\%$ across the four widths, an eight-fold improvement at the full width of one hundred qubits. The end-to-end Q-CTRL pipeline returned substantially higher exact-output success on these instances. Because the managed pipeline controls layout, compilation, scheduling, suppression, and measurement processing internally, the present experiment does not isolate the contribution of any individual component. Returning the exact 100-bit target in \(76.5\%\) of the shots is a notable full-width result within this execution campaign.

Measurement twirling, by contrast, provides no consistent improvement over raw execution. It adds $1.8$ points on the easiest Bernstein--Vazirani instance, subtracts a few points elsewhere, and at twenty counting qubits of phase estimation it returns $0.009\%$ against the raw $0.037\%$, a difference of a handful of shots. The null result has a clean explanation rather than being an anomaly. Twirling symmetrizes readout errors so that their bias can be removed from \emph{expectation values} in post-processing; it does not reduce the total readout error rate, and an exact-bitstring metric offers no averaging in which a symmetrized error could cancel. Redistributing the same error budget randomly across the register can even lower the probability of the single perfect outcome, which is precisely the small negative drift the table shows. Under the exact-output metric and measurement-twirling configuration tested here, twirling did not provide a consistent sampling improvement.

These gains come at essentially no premium in quantum-processor time. IBM jobs, batched by width, consumed between 10 and 36 QPU seconds each, and Q-CTRL jobs consumed 20 to 29 seconds per algorithm family on the three structured workloads. Within the granularity of these numbers, managed suppression is cost-neutral, which will matter for the synthesis in Section~\ref{sec:tradeoff}.

Table~\ref{tab:sampler_perqubit} shows that Q-CTRL maintains effective per-qubit rates between $98.14$ and $99.77\%$ for Bernstein--Vazirani, GHZ, and the random mirror circuits, while the IBM configurations range from $95.20$ to $98.35\%$. Phase estimation produces the clearest separation. Q-CTRL gradually declines from $97.54$ to $93.35\%$ as the width grows from $n=10$ to $n=30$, while IBM configurations fall from approximately $95.7\%$ to experimental zero, with no successful shots observed at $n=30$.

\begin{table}[!htb]
\centering
\caption{Effective per-qubit success rate $P^{1/n}$ in percent by algorithm, circuit width, and configuration. This aggregate metric normalizes the global success probability by width.}
\label{tab:sampler_perqubit}
\begin{tabular}{llrrr}
\toprule
Algorithm & $n$ & IBM raw & IBM twirl & Q-CTRL \\
\midrule
BV  & 25  & 97.54 & 97.67 & 99.10 \\
BV  & 50  & 95.48 & 95.20 & 98.14 \\
BV  & 75  & 95.54 & 95.36 & 98.61 \\
\midrule
QPE & 10  & 95.85 & 95.67 & 97.54 \\
QPE & 20  & 67.32 & 62.83 & 95.23 \\
QPE & 30  & 0.00  & 0.00  & 93.35 \\
\midrule
GHZ & 25  & 98.35 & 98.30 & 98.83 \\
GHZ & 50  & 98.12 & 97.92 & 98.44 \\
\midrule
RND & 25  & 98.20 & 98.08 & 99.73 \\
RND & 50  & 97.62 & 97.57 & 99.74 \\
RND & 75  & 97.74 & 97.60 & 99.77 \\
RND & 100 & 97.64 & 97.54 & 99.73 \\
\bottomrule
\end{tabular}
\end{table}

\FloatBarrier
\subsection{Estimator: accuracy against the exact reference}
\label{sec:estres}


Table~\ref{tab:estres} reports the expectation values, the provider-returned \texttt{stds} fields, the exact MPS references, and the corresponding absolute errors. The values following the $\pm$ sign are reproduced directly from each provider. Because Qiskit Estimator implementations do not enforce a common statistical definition for \texttt{stds}, these quantities are retained only for transparency and are not compared across providers or used to calculate $\Delta$, $S_O$, $S_H$, or the accuracy rankings. Absolute errors were calculated from the unrounded numerical values; the displayed expectation values and MPS references are rounded.

\begin{table}[!htb]
\centering
\caption{Estimator results for the chain-averaged observables of Eq.~\eqref{eq:obs}. Each entry lists the reported expectation value followed by the provider-returned \texttt{stds} field, the MPS reference, and the absolute error \(\Delta\). The \texttt{stds} fields have implementation-specific definitions and are not compared or used in any accuracy calculation. Bold marks the smallest error for each observable and size.}
\label{tab:estres}
\begin{tabular}{lllrrr}
\toprule
Obs. & $n$ & Configuration & $\langle O\rangle \pm \texttt{stds}$ & MPS ideal & $\Delta$ \\
\midrule
$X$  & 25 & IBM raw             & $0.4928 \pm 0.0010$ & 0.6606 & 0.1679 \\
     &    & IBM TREX + twirling & $0.5669 \pm 0.0017$ & 0.6606 & 0.0938 \\
     &    & Q-CTRL              & $0.6263 \pm 0.1557$ & 0.6606 & 0.0344 \\
     &    & QESEM               & $0.6799 \pm 0.0213$ & 0.6606 & \textbf{0.0193} \\
\addlinespace
$X$  & 50 & IBM raw             & $0.5062 \pm 0.0007$ & 0.6520 & 0.1458 \\
     &    & IBM TREX + twirling & $0.5560 \pm 0.0011$ & 0.6520 & 0.0960 \\
     &    & Q-CTRL              & $0.6101 \pm 0.1119$ & 0.6520 & 0.0419 \\
     &    & QESEM               & $0.6149 \pm 0.0172$ & 0.6520 & \textbf{0.0371} \\
\addlinespace
$X$  & 75 & IBM raw             & $0.5177 \pm 0.0005$ & 0.6491 & 0.1314 \\
     &    & IBM TREX + twirling & $0.5311 \pm 0.0010$ & 0.6491 & 0.1180 \\
     &    & Q-CTRL              & $0.6127 \pm 0.0912$ & 0.6491 & 0.0364 \\
     &    & QESEM               & $0.6220 \pm 0.0257$ & 0.6491 & \textbf{0.0271} \\
\midrule
$ZZ$ & 25 & IBM raw             & $0.3919 \pm 0.0010$ & 0.3887 & \textbf{0.0032} \\
     &    & IBM TREX + twirling & $0.3406 \pm 0.0019$ & 0.3887 & 0.0481 \\
     &    & Q-CTRL              & $0.3672 \pm 0.1898$ & 0.3887 & 0.0214 \\
     &    & QESEM               & $0.3960 \pm 0.0178$ & 0.3887 & 0.0073 \\
\addlinespace
$ZZ$ & 50 & IBM raw             & $0.3724 \pm 0.0007$ & 0.3910 & 0.0186 \\
     &    & IBM TREX + twirling & $0.3347 \pm 0.0013$ & 0.3910 & 0.0563 \\
     &    & Q-CTRL              & $0.3745 \pm 0.1324$ & 0.3910 & 0.0165 \\
     &    & QESEM               & $0.4046 \pm 0.0133$ & 0.3910 & \textbf{0.0136} \\
\addlinespace
$ZZ$ & 75 & IBM raw             & $0.3289 \pm 0.0006$ & 0.3917 & 0.0628 \\
     &    & IBM TREX + twirling & $0.3199 \pm 0.0010$ & 0.3917 & 0.0718 \\
     &    & Q-CTRL              & $0.3714 \pm 0.1079$ & 0.3917 & 0.0203 \\
     &    & QESEM               & $0.3834 \pm 0.0216$ & 0.3917 & \textbf{0.0083} \\
\bottomrule
\end{tabular}
\end{table}

The transverse magnetization is the canonical decoherence signal, and raw execution reads like the textbook. It underestimates $m_X$ by $0.17$, $0.15$, and $0.13$ at the three sizes, reporting for instance $0.4928$ against the ideal $0.6606$ at 25 qubits, , consistent with net attenuation of a coherence-sensitive observable toward zero. IBM's TREX-plus-twirling configuration recovers roughly half of that deficit at 25 and 50 qubits, cutting the error to $0.094$ and $0.096$, but its benefit erodes at 75 qubits, where the residual error increases to \(0.118\), indicating that the tested TREX-plus-twirling configuration does not recover the magnetization as effectively at this width. The two managed stacks change the character of the result rather than its degree: Q-CTRL holds the error between $0.034$ and $0.042$ and QESEM between $0.019$ and $0.037$ across all sizes, with no systematic growth from 25 to 75 qubits. Across the three widths, the magnetization MAE is \(0.1484\) for IBM raw, \(0.1026\) for IBM TREX + twirling, \(0.0376\) for Q-CTRL, and \(0.0278\) for QESEM. The managed stacks therefore produce substantial and width-consistent improvements in magnetization accuracy, with error reductions of approximately \(3.95\times\) for Q-CTRL and \(5.33\times\) for QESEM relative to raw execution.

\begin{figure}[!htbp]
\centering
\includegraphics[width=0.74\textwidth]{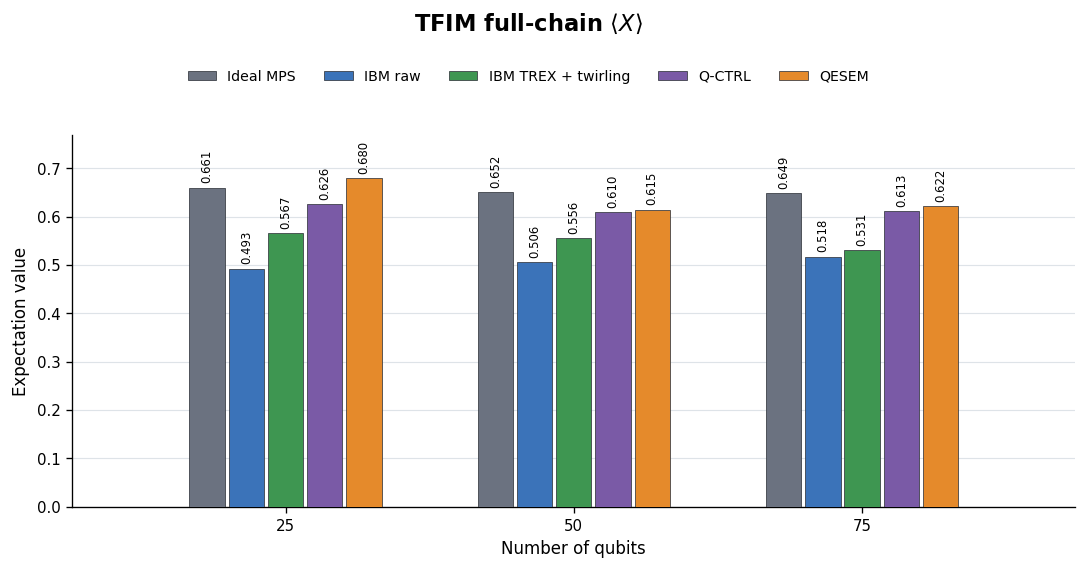}
\caption{Chain-averaged transverse magnetization $m_X$ by size and configuration, with the exact MPS reference shown in gray. Raw execution and, to a lesser degree, TREX-plus-twirling underestimate the observable, while the managed stacks track the reference at every size.}
\label{fig:xev}
\end{figure}

\begin{figure}[!htbp]
\centering
\includegraphics[width=0.65\textwidth]{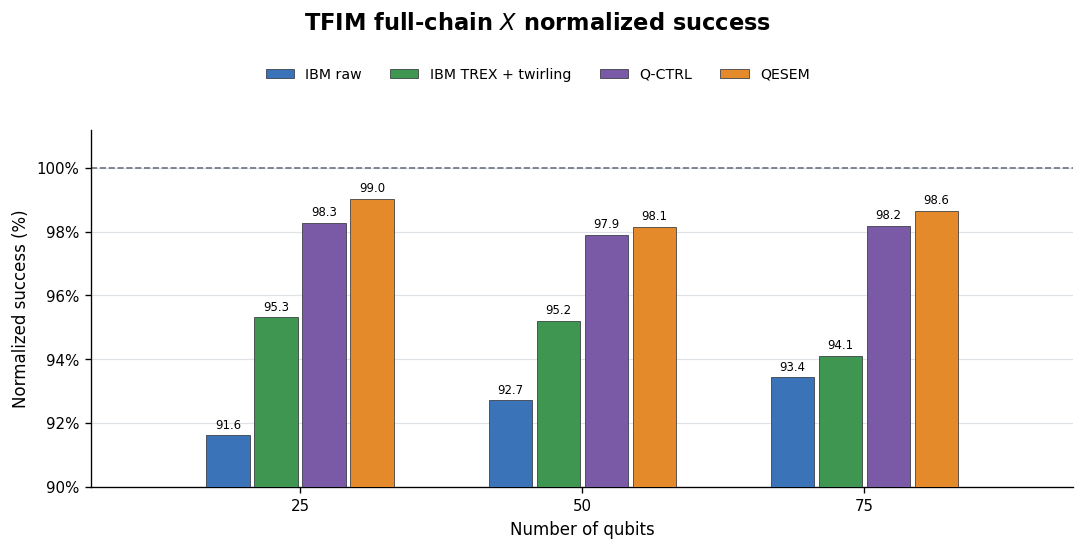}
\caption{Normalized success score of Eq.~\eqref{eq:score} for the transverse magnetization.}
\label{fig:xns}
\end{figure}

The tested IBM TREX-plus-twirling configuration returned a larger \(c_{ZZ}\) absolute error than raw execution at all three widths. The especially small raw error at \(n=25\) may reflect accidental cancellation among multiple hardware biases, but the present end-to-end data do not identify the underlying mechanism. The result therefore demonstrates that an error-management setting that improves one observable need not improve another measured from the same circuit. Q-CTRL and QESEM did not exhibit the same error ordering in these runs.

\begin{figure}[!htbp]
\centering
\includegraphics[width=0.65\textwidth]{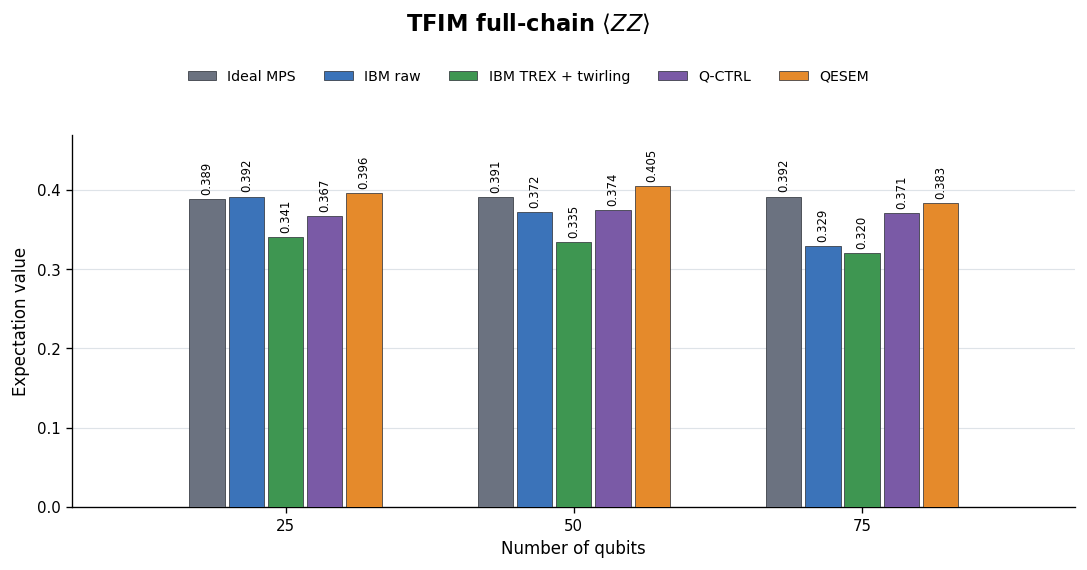}
\caption{Chain-averaged nearest-neighbor correlator \(c_{ZZ}\) by size and configuration, with the MPS reference shown in gray. The tested IBM TREX-plus-twirling configuration lies below the reference at every size, while the particularly small raw error at \(n=25\) may reflect accidental cancellation among hardware biases.}
\label{fig:zzev}
\end{figure}

\begin{figure}[!htbp]
\centering
\includegraphics[width=0.6\textwidth]{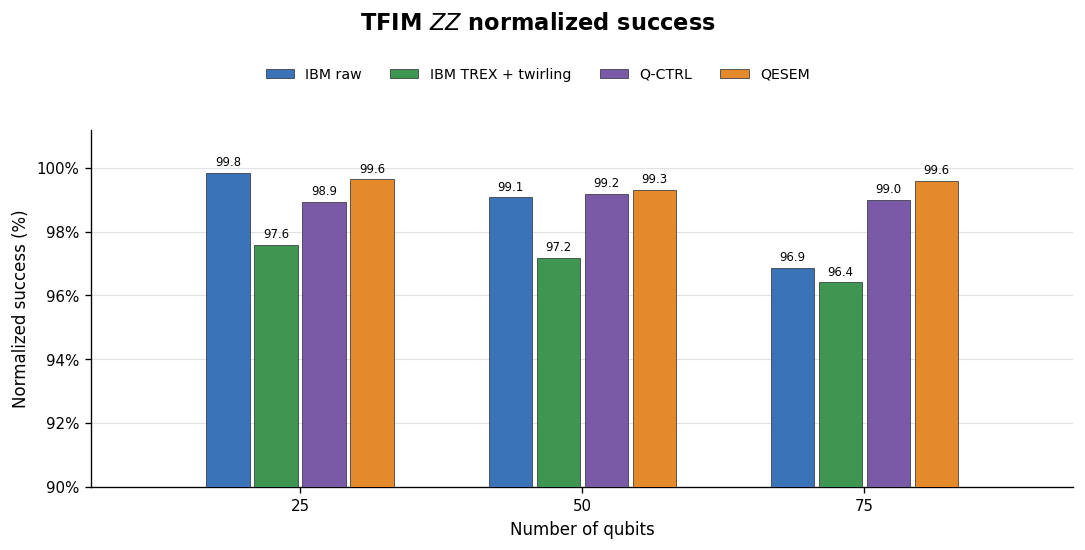}
\caption{Normalized success score of Eq.~\eqref{eq:score} for the nearest-neighbor correlator.}
\label{fig:zzns}
\end{figure}

To summarize the six MPS-referenced comparisons without obscuring the observable-specific behavior, Table~\ref{tab:estimator_mae} reports the mean absolute error separately for \(m_X\) and \(c_{ZZ}\), together with the overall MAE across all six observable--size cases. IBM raw execution has an overall MAE of \(0.0883\), while IBM TREX + twirling reduces it modestly to \(0.0807\). Q-CTRL and QESEM reduce the aggregate error to \(0.0285\) and \(0.0188\), respectively. Relative to raw execution, these correspond to error-reduction factors of \(1.09\), \(3.10\), and \(4.70\). The observable-specific values also preserve the correlator inversion visible in Table~6: TREX + twirling improves the mean magnetization error but increases the mean correlator error relative to raw execution.

\begin{table}[htbp]
    \centering
    \caption{Mean absolute error relative to the MPS reference.
    Observable-specific MAEs average over \(n=25\), \(50\), and \(75\);
    the overall MAE averages all six observable--size cases.}
    \label{tab:estimator_mae}
    \small
    \begin{tabular}{@{}lcccc@{}}
        \toprule
        Configuration
        & \(\mathrm{MAE}_{m_X}\)
        & \(\mathrm{MAE}_{c_{ZZ}}\)
        & Overall MAE
        & Reduction factor \\
        \midrule
        IBM raw
        & 0.1484
        & 0.0282
        & 0.0883
        & 1.00 \\

        IBM TREX + twirling
        & 0.1026
        & 0.0587
        & 0.0807
        & 1.09 \\

        Q-CTRL
        & 0.0376
        & 0.0194
        & 0.0285
        & 3.10 \\

        QESEM
        & \textbf{0.0278}
        & \textbf{0.0097}
        & \textbf{0.0188}
        & \textbf{4.70} \\
        \bottomrule
    \end{tabular}
\end{table}

The reduction factor is defined as \(\mathrm{MAE}_{\mathrm{IBM\,raw}}/\mathrm{MAE}_{p}\), so a value
greater than one indicates a reduction in aggregate error relative to raw execution.

\begin{figure}[!htbp]
\centering
\includegraphics[width=0.6\textwidth]{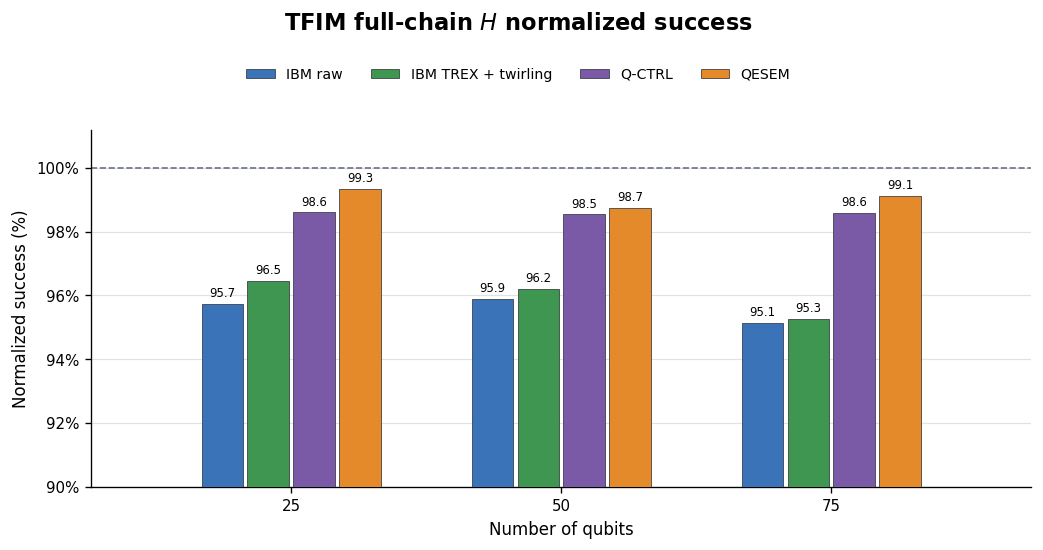}
\caption{Combined score $S_H = (S_X + S_{ZZ})/2$ by size and configuration. The managed stacks are separated from the built-in options by roughly three points, corresponding to an absolute-error difference of about $0.06$.}
\label{fig:hns}
\end{figure}

The normalized scores provide a complementary bounded representation of the same MPS-referenced absolute errors. Because Eq.~\eqref{eq:score} is a linear transformation of \(\Delta\), the normalized scores preserve the same aggregate ordering established by the MAE results in Table~\ref{tab:estimator_mae}. QESEM leads with $S_H$ between $98.7$ and $99.3$ across the three sizes, Q-CTRL follows closely at approximately $98.6$, and the two IBM configurations trail at $95.3$ to $96.5$ for TREX-plus-twirling and $95.1$ to $95.9$ for raw execution, the near-coincidence of the last two reflecting the correlator inversion that cancels the magnetization gain. Figure~\ref{fig:hns} makes the same point graphically: the managed stacks are separated from the built-in options by roughly three points on a scale where each point corresponds to an absolute error of $0.02$.

The provider-returned \texttt{stds} values span markedly different numerical scales, but these differences are not interpreted as differences in uncertainty quality. Qiskit Estimator implementations do not enforce a common statistical definition for this field, and each provider may construct it differently. We therefore retain the values as returned metadata for transparency, while all cross-provider conclusions are based on the reported expectation values, the common MPS references, the case-level absolute errors \(\Delta\), the aggregate MAEs of Table~\ref{tab:estimator_mae}, and the complementary normalized scores of Eq.~\eqref{eq:score}.

\FloatBarrier
\subsection{Accuracy and reported QPU-time usage}
\label{sec:tradeoff}

\begin{figure}[!htbp]
\centering
\includegraphics[width=0.72\textwidth]{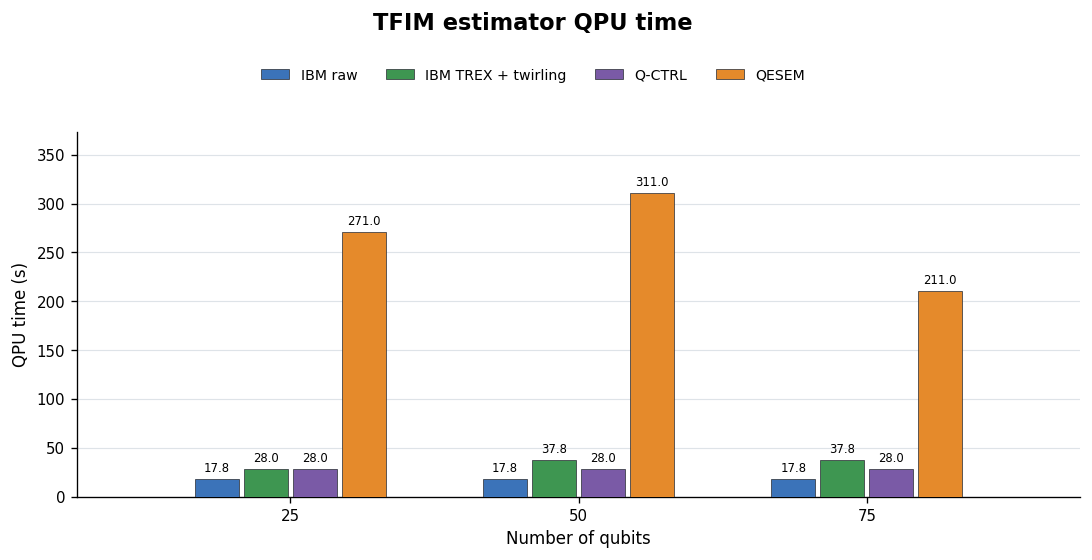}
\caption{Provider-reported QPU seconds per Estimator job by size and configuration. QESEM used \(7.5\)--\(11.1\times\) the QPU time of Q-CTRL while producing the lowest aggregate MAE. The tested IBM and Q-CTRL configurations remained within the same order of reported QPU time.}
\label{fig:eqpu}
\end{figure}

Figure~\ref{fig:eqpu} compares the provider-reported QPU seconds for the Estimator jobs. IBM raw execution used \(17.8\,\mathrm{s}\), IBM TREX plus twirling used \(28.0\)--\(37.8\,\mathrm{s}\), Q-CTRL used \(28.0\,\mathrm{s}\), and QESEM used \(211\)--\(311\,\mathrm{s}\). Relative to raw execution, Q-CTRL reduced the six-case MAE by a factor of \(3.10\) at QPU times within the same order as the tested IBM configurations. QESEM reduced the MAE by a factor of \(4.70\) while using \(7.5\)--\(11.1\times\) the reported QPU time of Q-CTRL. These measurements characterize provider-reported QPU usage only. Monetary price, third-party licensing, queueing delay, classical processing time, and end-to-end wall-clock latency were not evaluated.

On the Sampler workloads, Q-CTRL returned the highest success probability for every tested instance. For the three structured circuit families, its reported QPU times were within the same order as the IBM configurations. On the Estimator workloads, Q-CTRL and QESEM therefore occupy two distinct provider-native operating points in this execution campaign: Q-CTRL is the lower-QPU-time managed configuration, whereas QESEM gives the lowest aggregate error at higher QPU use. The tested IBM TREX-plus-twirling configuration provides a built-in baseline whose effect is observable dependent.

\FloatBarrier
\section{Conclusions}
\label{sec:conclusions}

We have reported a two-primitive benchmark of commercial error suppression and mitigation on a 156-qubit IBM Heron r3 processor. IBM and Q-CTRL configurations used matched requested budgets of \(2^{15}\) shots, while QESEM used provider-native precision and QPU-time controls. A seeded, manifest-driven workflow records the
logical circuits, configurations, and job identifiers used in the study. The design respects the fundamental boundary of the field, comparing suppression pipelines on sampling tasks, where bitstrings themselves must be protected, and the full suppression-plus-mitigation stack on estimation tasks, where an exact matrix-product-state reference allows accuracy to be measured rather than argued.

The findings are consistent across the two tracks and simple to act on. Within this execution campaign, Q-CTRL returned the highest exact-output or valid-set success probability for every tested Sampler instance. The largest differences occurred for quantum phase estimation and randomized mirror circuits. On the three structured circuit families, the reported QPU times remained within the same order as those of the IBM configurations; per-instance QPU times were not available for the mirror-circuit batch. Measurement twirling alone did not consistently improve exact-output success under the tested settings. This result applies to the provider-returned counts and exact-output metrics used here and should not be interpreted as a general assessment of distribution-level readout mitigation. For estimation, the managed stacks substantially reduced the aggregate MPS-referenced error, but at different QPU-time operating points. Across the six observable--size cases, the mean absolute error was \(0.0883\) for IBM raw execution, \(0.0807\) for IBM TREX + twirling, \(0.0285\) for Q-CTRL, and \(0.0188\) for QESEM. Thus, Q-CTRL and QESEM reduced the aggregate error by factors of \(3.10\) and \(4.70\), respectively, relative to raw execution.

These accuracy gains came with markedly different reported QPU-time requirements. Q-CTRL used \(28\) seconds per Estimator job, comparable to the \(17.8\)--\(37.8\) seconds used by the tested IBM configurations. QESEM achieved the lowest aggregate error while using \(211\)--\(311\) seconds per job, corresponding to \(7.5\)--\(11.1\) times the QPU time used by Q-CTRL. Compared directly, QESEM reduced the aggregate MAE by \(34.1\%\) relative to Q-CTRL while consuming substantially more QPU time. The Estimator results therefore identify two distinct accuracy--resource operating points: Q-CTRL provides the lower-QPU-time managed option, whereas QESEM provides the lowest observed error at higher QPU use. IBM TREX + twirling produced only a modest aggregate improvement because its reduction in magnetization error was partially offset by an increase in correlator error, demonstrating that mitigation performance can depend on the measured observable. The provider-returned \texttt{stds} fields are retained for transparency but are not included in this comparison because their statistical definitions differ across implementations.

Several limitations bound the generality of these conclusions. Each provider instance condition was represented by one execution job, and each randomized circuit family used one seeded circuit instance per
width. The shot-level uncertainty therefore quantifies finite sampling within a job, not variation across independent calibration windows or circuit instances. The IBM zero-noise-extrapolation arm was configured but not executed within the study's compute budget. The comparison therefore includes the tested IBM resilience-level-2 configuration with explicitly enabled gate twirling, but does not include IBM ZNE or PEC. QESEM could not be evaluated on sampling because no such interface exists, Q-CTRL's QPU times for the mirror-circuit batch were not reported by the platform, and a fuller study would add repeated runs across calibration cycles to convert single measurements into distributions. Within this execution campaign, the results support two practical observations: managed suppression produced the highest success probabilities on the tested sampling workloads, while Q-CTRL and QESEM occupied distinct Estimator accuracy - QPU-time operating points, with IBM's built-in configurations providing an accessible but observable-dependent baseline.

As error correction matures, the division of labor benchmarked here will shift, but it will not disappear. Early fault-tolerant machines are widely expected to combine logical qubits with mitigation on top~\cite{cai2023,aharonov2025qesem}, and independent, like-for-like measurements of what each software layer contributes will remain the only reliable guide to a fast-moving commercial landscape. We hope the workload suite and the submit-and-collect protocol introduced here provide a reusable template for such measurements.

\FloatBarrier
\subsection*{Data availability}
The benchmark notebooks, the JSON job manifests, the seeds, and the scripts that regenerate every figure and table from job-level records are available from the corresponding author on reasonable request.

\subsection*{Acknowledgments}

The authors acknowledge BasQ--Basque Quantum for facilitating access to the quantum-computing resources and Qiskit Functions used in this study, and the eVIDA Research Group at the University of Deusto for promoting
the collaboration that led to this work. The authors also thank Yuval Baum, for technical guidance concerning the use of the Q-CTRL Performance Management service.

\subsection*{Competing interests}

D.S. is a paid consultant to Q-CTRL. This consultancy is separate from the present study and provided no financial support for the work. The authors retained full control over the benchmark design, workload and metric selection, execution, access to the data, data analysis, interpretation of the results, preparation of the manuscript, and the decision to submit the work for publication. Q-CTRL provided technical advice concerning the use of the Q-CTRL service but did not participate in metric selection, data analysis, interpretation of the results, or the decision to publish. The remaining authors declare no competing interests.

\end{document}